\documentclass[twocolumn,twoside,slac]{revtex4}
\usepackage{graphicx}
\usepackage{fancyhdr}
\begin{document}

\title{Experience with the Open Source based implementation for ATLAS
Conditions Data Management System}

\author{A.Amorim}

\author{J.Lima}

\author{C.Oliveira}

\author{L.Pedro}

\author{N.Barros}
\affiliation{CFNUL/FCUL, Universidade de Lisboa, Portugal}

\begin{abstract}
Conditions Data in high energy physics experiments is frequently seen
as every data needed for reconstruction besides the event data
itself. This includes all sorts of slowly evolving data like detector
alignment, calibration and robustness, and data from detector control
system. Also, every Conditions Data Object is associated with a time
interval of validity and a version. Besides that, quite often is
useful to tag collections of Conditions Data Objects altogether. These
issues have already been investigated and a data model has been
proposed and used for different implementations based in commercial
DBMSs, both at CERN and for the BaBar experiment. The special case of
the ATLAS complex trigger that requires online access to calibration
and alignment data poses new challenges that have to be met using a
flexible and customizable solution more in the line of Open Source
components. Motivated by the ATLAS challenges we have developed an
alternative implementation, based in an Open RDBMS. Several issues
were investigated and will be described in this paper:

- The best way to map the conditions data model into the relational
 database concept considering what are foreseen as the most frequent
 queries.

- The clustering model best suited to address the scalability problem.

- Extensive tests were performed and will be described.

The very promising results from these tests are attracting the
attention from the HEP community and driving further developments.

\end{abstract}

\maketitle

\thispagestyle{fancy}


\section{Introduction}

The ATLAS Conditions Database Management System, ATLAS ConditionsDB
for short, is the system responsible for storing data from conditions
associated with ATLAS experiment. Conditions Data reflect the
conditions in which the experiment was performed and the actual
physics data was taken. This includes detector calibration and
alignment, robustness data from Detector Control System (DCS) and
detector description. Stating in another way, conditions data is the
slowly evolving data associated with the experiment besides event 
data itself. This data is produced in many different subsystems and 
each provides data at different rate and store objects of different 
granularities. From the data consumer side the situation is similar, 
having consumers with many different demands concerning data rate and
latency.

The Open Source based Implementation of the ATLAS ConditionsDB
described in this paper is the result of a development effort of
almost 2 years. The aim of this effort was to provide an alternative
solution to those that have been presented by the CERN-IT, based in
commercial solutions like Objectivity and Oracle. Several reasons leaded us
to select an Open Source Solution. Namely: 

\begin{itemize}
\item It's free of charge.
\item A growing community of programmers have a good understanding of open
  source technologies.
\item Code availability allow fine tuning and specific optimizations
  when necessary.
\item available for most common platforms. 
\end{itemize}

Being free of charge, by itself, is not a big issue when one take into
consideration the huge budgets involved in particle physics
experiments, but it has been crucial to widespread these technologies
amongst the communities of programmers.
The three remaining items are key ingredients to decouple the
applications from the hardware or software vendors, as required for
long term projects. 
Furthermore it is the authors belief that is possible and preferable to 
achieve the desired performance by carefully taken design choices than by 
relying in the technological advanced features present in high-end 
commercial solutions.

This implementation uses a Relational Database Management System (RDBMS)
server as the underlying storage technology. The development was done 
using mostly MySQL, but any other RDBMS that understands standard SQL
could be used. Yet the choice of MySQL was sensible. The MySQL is proven
to be a reliable and very fast RDBMS, although it lacks some important 
features found in many common commercial RDBMS.

\section {The ATLAS Conditions Data Model}

In this section shall briefly explain what is the data model used
in the ATLAS ConditionsDB. In this context, data model means what
model is exposed to the user/client through the API \cite{specs-ref}.
This is something different from the model used to actually store the
data. The later shall be designate \emph{database schema} and will be
explained bellow. The API ConditionsDB specification \cite{specs-ref}
was the starting point for this work.

\begin{figure*}[t]
\centering
\includegraphics[width=135mm]{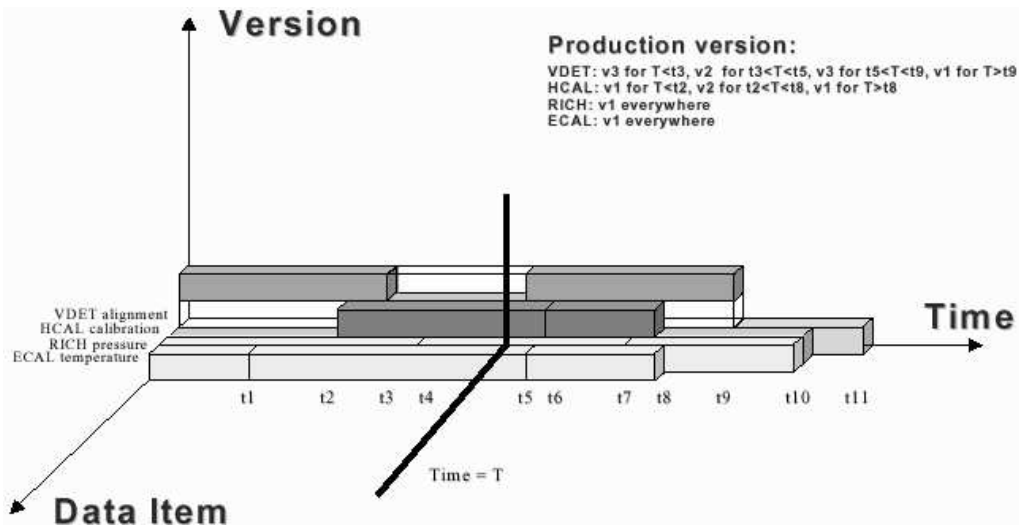}
\caption{Conditions Data variation axis.} \label{axis-fig}
\end{figure*}

The present ATLAS ConditionsDB model shares many of the features found
in the BaBar ConditionsDB model \cite{babar-ref}. The data is organized
in a filesystem like hierarchical structure, where each folder holds a
particular type of data. A version and a time of validity (IOV) is
also associated with the data. Figure \ref{axis-fig} shows the three
variation axis associated with the data. 
No matter which underlying technology is used, the exposed model is
always the same. In this implementation, a relational model was used. 
The first problem was understanding how to map this data model that 
uses an hierarchical folder structure on the underlying relational model 
which uses tables and columns. The second problem was understanding the
type of queries that were being used: which were the most common ones; 
how could one redesign the tables to make the life easier for the 
database server.

\section{Design Aspects}

In this section we will try to give an overview of the architecture
and design of the Open Source ATLAS ConditionsDB Management System.
The system was designed with three features in mind: backend
independency; data volume scalability and extensibility.
By backend independency we mean that it should be possible, and easy,
to replace the underlying RDBMS, as long as it understands standard
SQL. Note that we carefully stated backend independency, not
technology independency, which has a more generic mean.
Data volume scalability should allow us to reach, at a certain point,
a $O(1)$ performance as function of data volume. It also means that
should be easy to switch the data, in small chunks, back and forth
from tertiary data storage.
Extensibility means that the software system must be modular with well
defined interfaces in order to improve maintainability and extension
of functionality with new modules. Next we will show how we
achieved each of these features.

\subsection{Data volume scalability}

This issue is closed connected to the overall performance of the
system, Nevertheless, it should be stressed that no post design
optimizations can help to get rid of a O(n!) behavior. This is an
obvious fact, although so often completely ignored. Yet our approach
to this problem was not to devise a full O(1) compliant design from
the beginning. We took what we can call a successive approximation
with design/test/redesign approach. The drawback of this approach is
that one takes a lot of effort in design, knowing from the beginning
that it will be thrown away. Lets be honest with ourselves, we end up
throwing away a lot of work even when we do not assume that fact from
the beginning.

The task of making a Database management system that can scale well to
the data volumes expected from an experiment like ATLAS pushes the
limits of our imagination. Every thing that can be done to improve
performance must be done. Not only the database schema must be
optimum for the problem in hands, one must understand which data 
clustering model is best suited for the Conditions Data.
Our implementation supports a data clustering model which should scale
very smoothly over large data volumes. By design the clustering can
naturally be made in a per folder basis. Moreover, our implementation
supports data partitioning inside folders in a time or version basis.
Thus, the data partitioning mechanism is highly configurable and
flexible enough to adapt be adapted to several situations.

\subsection{Backend independency}

Willing to use Open Source also means willing to get rid of supplier's
dependence. Using standards instead of proprietary or specific vendor
features is a good thing, anyway. Unless there was a very good reason
to stick with a specific backend technology e would always prefer to
be able to switch to a different backend whenever we feel
like. For the time being almost all the development has been done with
the MySQL backend, but plans exist to test our implementation with
other RDBMS backends and even two support different backends in the
same implementation. A successful experiment have already been made
that used the Postgress as a RDBMS backend.

\subsection{Extensibility}

\begin{figure*}[t]
\centering
\includegraphics[width=135mm]{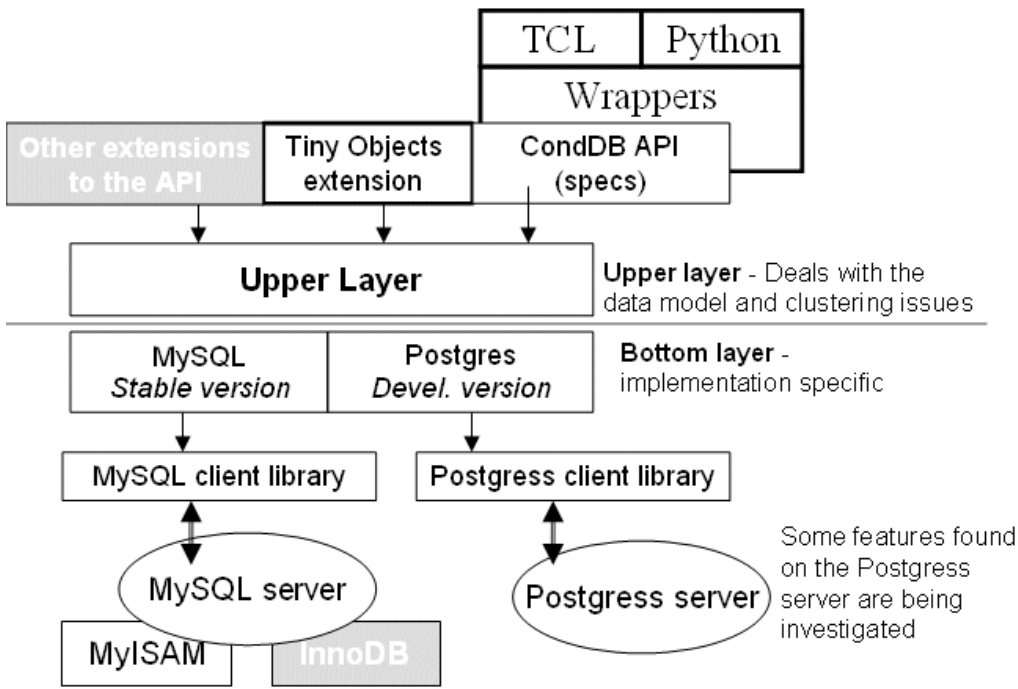}
\caption{System software layout.} \label{modul-fig}
\end{figure*}

A modular design policy was implemented since the beginning. The
software is composed of several layers and each accomplishes a
particular task. This design has made possible, or at least easier to
implement new features and to maintain the code in a cooperative
development environment. Figure \ref{modul-fig} depicts the modular 
software architecture as it is for the latest version.

The recent developments that will be described in this paper on
section \ref{new-sec}.

\section{New Features\label{new-sec}}

\subsection{New database schema}

\begin{figure*}[t]
\centering
\includegraphics[width=135mm]{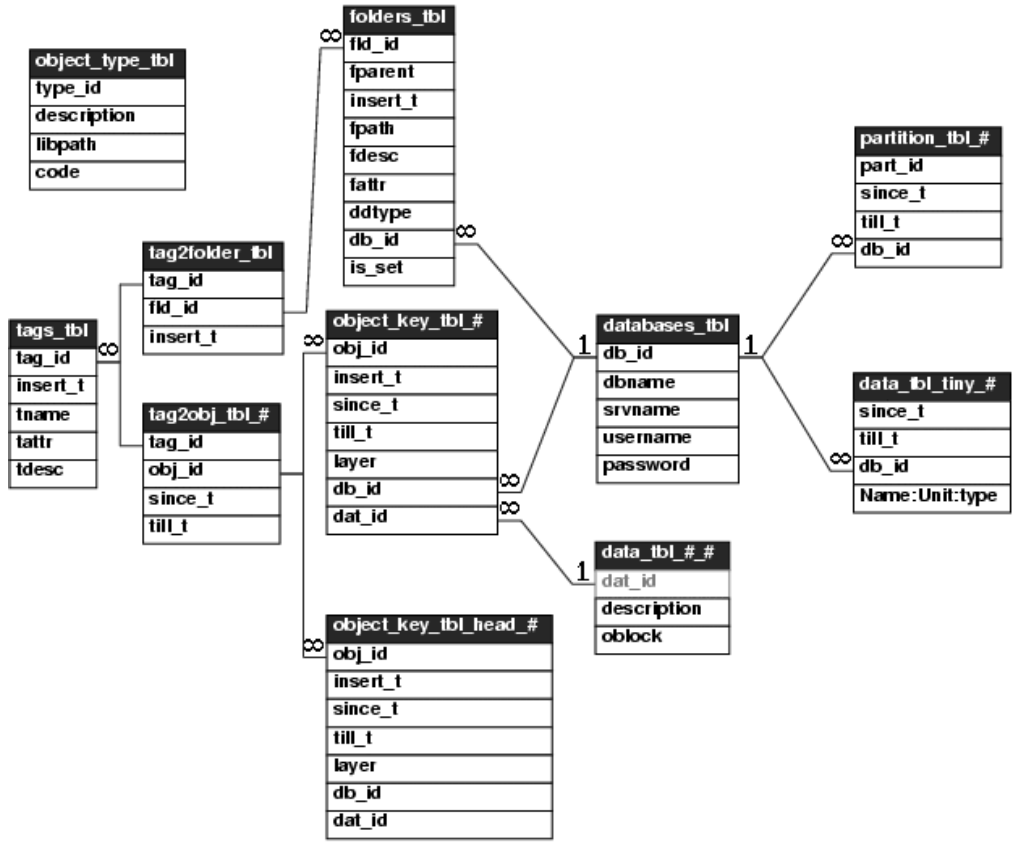}
\caption{The new database schema.} \label{schema-fig}
\end{figure*}

The new database schema was designed in order to allow
extension of the API to support new features. Notably the Tiny
Objects. A considerable redesign of the database schema was also
needed in order to improve storage performance. The results from
comparison tests before and after redesign are presented in section 
\ref{tests-sec}. Figure \ref{schema-fig} shows the table schema as it
was at time or writing this paper.

\subsection{Support for tiny objects}

The \emph{Tiny Objects} in the context of the ConditionsDB were proposed few time
ago to the CERN ATLAS community in response to a requirements expressed
from users whom wished to store objects with simple structures but
with a huge granularity. The best example of objects of this kind are
the data coming from the DCS part of the detector. To deal with this type of
objects the API was extended in order to handle the \texttt{PVSS}
interface.

What the people from DCS needed was a way to efficiently store and retrieve 
very small data objects, typically, integers or floats. The idea
of \emph{Tiny Objects} extension is to provide such way at the cost of  
dropping some general features. Features which, after all, are not need
in this case.

The DCS objects are often simple scalar object representing the ddp or the 
temperature in certain detector elements. This are very small object that 
will get a huge storage overhead if stored with the primitive ConditionsDB
approach. On the other hand, people from the DCS don't foresee the
absolute need for using versions for their objects, only the interval
of validity.

Starting from version 0.3, the API features storage and retrieving of tiny
objects. That feature actually match a user requirement.
Remarkable is the fact that people from the DCS are already using this
brand new \emph{Tiny Objects} extension in the test beam at CERN.

\subsection{TCL Test Framework}

The first tests performed on this particular implementation of the
ATLAS Conditions Database were based in the test suite from the Oracle
based IT's implementation. The main purpose was then to evaluate in
what extent the behavior of the API was the same regardless of the
underlying implementation. Yet these tests were inappropriate for an
optimization phase: they do not fully explore the eventual weaknesses
and bottlenecks of the implementation and, on the other end, they do
not reproduce the foreseen runtime operations. Thus, recently, a
considerable effort was devoted to fully understand the system in
terms of performance. Soon become clear that we needed a flexible,
highly configurable, test framework. More like a scripting system. So,
wrappers for the CondDB's C++ API classes were developed for TCL,
giving rise to a complete CondDB scripting system.

To illustrate the idea we reproduce a simple test program, that opens
a database and create a folder structure, both in C++ and in
TCL. Immediately becomes clear that the TCL version is more concise,
besides being less error prone and having the advantage of not
requiring the usual compile/link cycle.

The exception handling code was removed from both, the C++ example and
the TCL example for sake of clarity.
 
\paragraph{C++ example}

{\tiny\begin{verbatim}
// File: "basicSession.cxx"
//
// Created at Mon Aug 26 18:25:22 WEST 2002
// by Jorge Lima, portuguese TDAQ group, 
// to the atlas collaborations.
//
// Based on the CERN IT Objectivity Implementation.
//

#include <ICondDBMgr.h>
#include <CondDBMySQLMgrFactory.h>

#include <string>
#include <iostream>
using namespace std;

int main ( int argc, char* argv[] )
{
    ICondDBMgr* CondDBmgr = 
         CondDBMySQLMgrFactory::createCondDBMgr();

    CondDBmgr->init();

    // Create a ConditionsDB

    CondDBmgr->startUpdate();
    CondDBmgr->createCondDB();
    CondDBmgr->commit();
        
    CondDBMySQLMgrFactory::destroyCondDBMgr( CondDBmgr );
    return 0;
}
\end{verbatim}}

\paragraph{TCL example}

{\tiny\begin{verbatim}
#!/bin/sh
#
# TCL Wrapper for ConditionsDB
# Lisbon ATLAS-DAQ Collaboration
# Ported from C++ to TCL by Jorge Lima 2003/04/15
#\
exec tclsh "$0" "$@"

package require conddb 1.0
 
set CondDBmgr \
     [CondDBMySQLMgrFactory::createCondDBMgr]

$CondDBmgr init

# Create a ConditionsDB

$CondDBmgr startUpdate
$CondDBmgr createCondDB
$CondDBmgr commit

CondDBMySQLMgrFactory::destroyCondDBMgr $CondDBmgr
\end{verbatim}}

\subsection{Other uses for scripting}

\begin{figure*}[t]
\centering
\includegraphics[width=135mm]{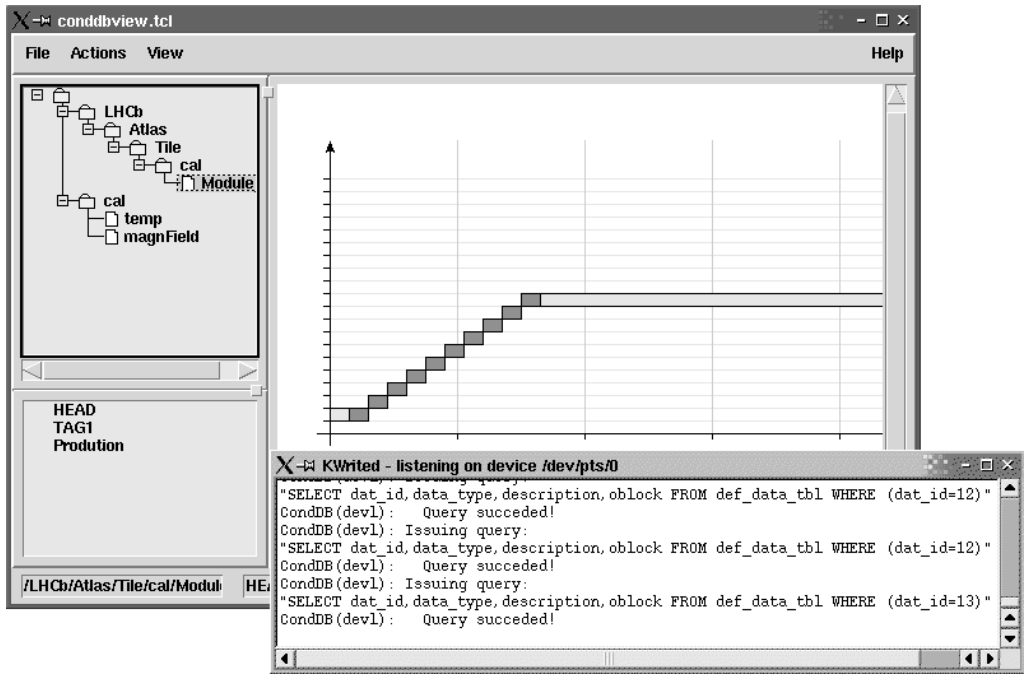}
\caption{Tcl/Tk browser for the ConditionsDB.} \label{browser-fig}
\end{figure*}

Although initialy developed as a test framework, the scripting bind to
the ConditionsDB has a very broad scope.  It is important to remark
that no performance penalty is noticeable by using the scripting
interface to the conditions DB instead of C++. Of course, the
scripting environment is not intended to replace the C++ programming
environment, and as more complex object manipulations are necessary
the scripting environment might became rather limited. Yet as far as
storing, retrieving and browsing is concerned a TCL program is as fast
as a C++ program.
For example the ConditionsDB TCL/TK browser, which used to interact
with the ConditionsDB through a command line based tool written in
C++, can now be completely rewritten in a very straight forward way.
For very long that users made suggestions regarding this browser in
order to make of it a more useful tool. Work is in progress in this
area and it will hopefully possible to provide very soon a complete
and fully functional browser using the TCL API. Figure \ref{browser-fig}

Moreover, the experience acquired while building this scripting
environment will make the job easier if we happen to need to write
bindings for other scripting language like, for instance, \emph{python} or \emph{php}.  

The TCL API also includes tools for performance measurement, which
make this API specialy suited for building performance tests.

\section{Performance Test Results\label{tests-sec}}

From time to time we have presented performance comparisons between
our implementation of the ATLAS ConditionsDB and the reference ORACLE
based implementation supported by the CERN-IT. Soon it became clear
that this was a development that was worth being investigated
further. Even before any optimizations or use of indexes, our
implementation with the MySQL backend, outperformed the ORACLE's one
by a factor of 10 to 50 for most common operations, as shown in table
\ref{oracle-comp-tbl}.

\begin{table*}[t]
\begin{center}
\caption{Comparisons between MySQL's and Oracle's implementations.}
\begin{tabular}{|r|r|r|r|r|}
\hline & \textbf{Oracle (local)} & \textbf{MySQL (local)} & 
         \textbf{Oracle (remote)} & \textbf{MySQL (remote)} \\
\hline \textbf{createFolderx} & 0m15.173s & 0m0.034s & 0m11.857s & 0m0.072s \\
\hline \textbf{storeDatax 10}
         & 0m4.973s & 0m1.447s & 0m5.434s & 0m1.127s \\
\hline \textbf{storeDatax 100}
         & 0m9.749s & 0m1.368s & 0m15.820s & 0m2.345s \\
\hline \textbf{storeDatax 10.000}
         & 9m22.103s & 0m23.175s & 11m12.554s & 0m56.929s \\
\hline \textbf{storeDatax 100.000}
         & 109m40.878s & 3m49.184s & 104m13.283s & 8m16.510s \\
\hline \textbf{storeDatax 1.000.000}
         & & 23m12.563s & & 40m48.256s \\
\hline \textbf{readDatax 10}
         & 0m0.324s & 0m0.025s & 0m0.955s & 0m0.058s \\
\hline \textbf{readDatax 100}
         & 0m1.403s & 0m0.050s & 0m3.061s & 0m0.135s \\
\hline \textbf{readDatax 10.000}
         & 2m1.919s & 0m2.851s & 4m37.315s & 0m7.926s \\
\hline \textbf{readDatax 100.000} 
         & 25m46.423s & 0m27.273s & 46m53.846s & 1m18.850s \\
\hline \textbf{readDatax 1.000.000}
         & & 4m40.315s & & 10m26.124s \\
\hline
\end{tabular}
\label{oracle-comp-tbl}
\end{center}
\end{table*}

There was still a problem to be solved: we identified a worst case for
storage operations where the performance was unusualy poor. Moreover,
storing an object took $O(n)$ time where n were the number of objects
already stored in the table.
Although it's clear that, in the ConditionsDB a storing is complex operation
(involving far more SQL queries) than a fetch operation, there was no
trivial explanation for such a different behavior between this case and
the storage in other situations. This test was not performed in
Oracle9i based implementation but, after investigating the problem, we
believe this was a problem specific to our implementation.

The problem was successful identified and corrected. Now the store
time is nearly constant with respect to the number of stored
objects. Though, this implied, not just simple optimizations, like
indexes or caching policies, but rather a schema redesign. The system is
converging to a $O(1)$ performance. There are, however, some operations
that still must be tuned. We can even expect complete table schema
redesign in order to cope completely with the $O(1)$ goal.

Current performance comparisons between our ConditionsDB
implementation, prior and after the schema redesign are shown in
tables 1 and 2. It must be stressed that the new implementation also
includes the extensions mentioned in section X. So, the new
implementation is, both, more functional and is closer to the $O(1)$
goal.

In the table for different results are shown, two for store operations
an two for read operations. The \emph{Store A} test stores objects having the
same interval of validity while the \emph{Store B} test stores objects
with overlapping intervals of validity, with monotonically increasing
start times. This was identified as a worst case situation for the store
operation using the old database schema. The \emph{Read A} and
\emph{Read B} tests iterate over the objects stored with the
\emph{store A} and \emph{store B} tests respectively.

\begin{table}[t]
\begin{center}
\caption{Performance test results with old table schema}
\begin{tabular}{|r|r|r|r|r|}
\hline \textbf{N. obj}  & \textbf{Store A} & \textbf{Read A} &
       \textbf{Store B} & \textbf{Read B}   \\
\hline     10 &  0m0.030s &  0m0.007s &    0m0.036s &  0m0.010s \\
\hline    100 &  0m0.277s &  0m0.134s &    0m0.851s &  0m0.219s \\
\hline   1000 &  0m5.393s &  0m0.784s &    0m9.428s &  0m0.772s \\
\hline  10000 & 0m25.127s &  0m3.409s &   10m7.253s &  0m7.694s \\
\hline  50000 & 2m14.046s & 0m17.873s & 4h3m51.000s & 0m35.907s \\
\hline
\end{tabular}
\label{old-schema-tbl}
\end{center}
\end{table}

\begin{table}[t]
\begin{center}
\caption{Performance test results with new table schema}
\begin{tabular}{|r|r|r|r|r|}
\hline \textbf{N. obj}  & \textbf{Store A} & \textbf{Read A} &
       \textbf{Store B} & \textbf{Read B}   \\
\hline     10 &  0m0.037s &  0m0.008s &   0m0.035s &  0m0.010s \\
\hline    100 &  0m0.196s &  0m0.031s &   0m0.318s &  0m0.062s \\
\hline   1000 &  0m2.180s &  0m0.263s &   0m3.608s &  0m0.651s \\
\hline  10000 & 0m22.151s &  0m2.572s &  0m44.693s &  0m6.285s \\
\hline  50000 & 1m55.495s & 0m14.725s &  3m50.663s & 0m33.950s \\
\hline
\end{tabular}
\label{new-schema-tbl}
\end{center}
\end{table}

\section{Future developments}

The ConditionsDB is a rapidly evolving subject. In the last months, not only
considerable human resources have been allocated to this development
effort which resulted in many improvements in several areas, but also,
the HEP community is getting acquired of the true dimension of the
problem and has started to contribute with their feedback.
Thus, it's not an easy task to foresee what will be the future
developments in this area. Nevertheless the next few modifications will most
likely to happen.

\subsection{New tagging mechanism}

The redesign o the tagging mechanism will be the next milestone. As
pointed out several times [Malon], the current tagging mechanism is,
both, incomplete and unreliable. This is not a problem specific to our
implementation. It is also present in the ORACLE's implementation as
it was in the Objectivity's one. The problem lies in the original API
specification [CondDBSPec] that is dated back to XX-XX-1999. Changing
this situation would require that we rewrite the API specification.

It is incomplete because doesn't allow tagging of other objects
besides head. This is contrary to what is stated in the requirements
collected so far [Req]. Several scenarios for tagging were presented
by [Malon] that cannot simply be accomplished by the current
implementation.

It is also unreliable, because it can effectively lead to data loss,
as no other way to reference a particular version exists besides
referencing it by head or by a tag. If after inserting some objects
one user wishes to tag the head for later reference, he/she can end up
tagging a completely different version, because some other user, in
the meantime, inserted new object, completely replacing the head.

A proper resolution to this problem necessarily passes by rewriting
the API specification with the involvement of all the community. Yet,
although many times stressed the importance of this question, no
significant advances have been seen from that side. A much ruder
approach will be to develop a working solution, get the feedback from
users, iterate a few times and finally, if we happen to converge to a
viable solution, write the specification. The later approach will be
taken for practical reasons.

\subsection{Expandable Objects}

The concept of \emph{Expandable object} will be soon
introduced. Briefly this are objects for which schema is known at the DBMS
level. This means that any client can query the DBMS about the object
structure and get as reply a platform independent representation.

The implementation for \emph{Expandable Objects} will benefit from the
work done for the \emph{Tiny Object}

In the present situation two main frameworks for Memory Databases exist in
the ATLAS context. Each use their own object schema. The Online Memory
Database is based in the OKS, and the Offline Memory Database is based
in the Athena Transient Store.
Typically, the Memory Databases will interact with the ConditionsDB
storing their objects as BLOBs. Anybody from outside a given framework
will be completely unable to understand objects stored from within
that framework. 

The \emph{Expandable Objects} concept will allow any client to browse through
the object schema because the database management system will be
schema aware.

\subsection{Minor issues}

Work has still to be done in many different areas. Now that most of
the performance bottlenecks have been removed, more realistic tests
involving many clients can be performed. Different multi-server
configurations can be experimented in order to understand in which
direction to go.

For a long time that an $<XML>$ integration in the API as been deferred
due to other priorities. An embedded $<XML>$ parser can be an effective
way of configuring the database in a way that is very extendible.

\section{Conclusions}

The ATLAS effort to manage the Conditions Database is extremely
demanding and complex. The system must cope with many different kinds
of data that must be accessible and distributed both online and
offline making it crucial to investigate carefully the performance
implications of each architecture option.

For large high energy physics experiments that have to run for a large
number of years it is mandatory to achieve decoupling from the
applications to the specific external software that is used. The
relational model with the recent object extensions provides the bases
of such independence. In the case o MySQL some object features already
present in other RDBMS, like the support for variable length arrays,
had to be implemented in a thin software layer on top of the basis
API.

The usage of Open Source relational DBMS has proven to be quite
adequate since it not only allows the easy portability to the
different platforms used online but also shows very good performance
results. Furthermore:

\begin{itemize}
\item The proposed and implemented clustering model, which can use
  multiple databases and servers, should scale over data volume and time.
\item Porting between different backend implementations doesn't
  constitute a major effort as verified while porting the original
  implementation to Postgress.
\item The MySQL implementation outperformed the Oracle's one by a
  factor of 50 on most usual operations.
\end{itemize}

There is also a strong effort to make the system more robust and
easy to use. The Open Source nature of the project and of all used
components makes it possible, for a wide variety of users, to run it
and provide valuable information to improve its robustness.

\begin{acknowledgments}

The authors wish to thank all the ATLAS-TDAQ community by their feedback and
valuable suggestions. A special word of thanks for David Malon for the
way he clearly pointed out some important design flaws.
Thanks to those contributing actively to the T-DAQ ConditionsDB
requirements \cite{webreq-ref} effort.

\end{acknowledgments}


\end{document}